\documentclass{ws-ijmpa}

\usepackage[super]{cite}
\usepackage{xcolor}
\usepackage[verbose,hypertexnames=false]{hyperref}
\hypersetup{colorlinks=false,allbordercolors=blue,pdfborderstyle={/S/U/W 1}}
\usepackage{pdflscape}
\usepackage{amsmath}
\usepackage{amsfonts}
\usepackage{amssymb}
\usepackage{graphicx}
\usepackage[titletoc]{appendix}
\usepackage{color}
\usepackage{hyperref}
\usepackage{cleveref}
\usepackage[rightcaption]{sidecap}
\usepackage{comment}
\usepackage{soul}
\usepackage{cancel}

\usepackage{dcolumn}

\usepackage{longtable}
\usepackage{floatrow}
\usepackage{array}
\usepackage{ctable}
\usepackage{multirow}
\usepackage{siunitx}
\usepackage{tabularx}
\usepackage{booktabs}
\usepackage{supertabular}

\graphicspath{{Graphics/}}

\def\be{\begin{equation}}
\def\ee{\end{equation}}
\def\bea{\begin{eqnarray}}
\def\eea{\end{eqnarray}}

\definecolor{vividviolet}{rgb}{0.62, 0.0, 1.0}
\definecolor{amaranth}{rgb}{0.9, 0.17, 0.31}
\definecolor{palatinateblue}{rgb}{0.15, 0.23, 0.89}
\definecolor{brightpink}{rgb}{1.0, 0.0, 0.5}
\definecolor{cornflowerblue}{rgb}{0.39, 0.58, 0.93}
\definecolor{deepcarminepink}{rgb}{0.94, 0.19, 0.22}
\definecolor{radicalred}{rgb}{1.0, 0.21, 0.37}

\hypersetup{ linktoc=all,
    colorlinks, linkcolor={palatinateblue},
    citecolor={brightpink}, urlcolor={amaranth}
}

\begin{document}

\markboth{Authors' Names}{Instructions for typing manuscripts (paper's title)}

\catchline{}{}{}{}{}
%
%%%%%%%%%%%%%%%%%%%%%%%%%%%%%%%%%%%%%%%%%%%%%%%%%%%%%%%%%%%%%%%%%%%%

\title{Investigating cosmic distance duality and dark energy evolution through intermediate and high-$z$ probes}

\author{Anna Chiara Alfano
}

\address{Scuola Superiore Meridionale, Largo S. Marcellino 10, 80138 Napoli, Italy.\\
a.alfano@ssmeridionale.it}

\maketitle

%\begin{history}
%\received{Day Month Year}
%\revised{Day Month Year}
%\accepted{Day Month Year}
%\published{Day Month Year}
%\end{history}

\begin{abstract}

We investigate deviations from the cosmic distance duality relation adopting model-dependent and -independent approaches using i) a Taylor expansion, ii) a power-law parameterization, iii) a logarithmic correction, iv) a $(2;1)$ Padé polynomial and v) a second order Chebyshev parameterization. We derive constraints on all parameters using observational Hubble data, galaxy clusters, type Ia supernovae, DESI data and gamma-ray bursts. Through Monte-Carlo Markov chain analyses adopting the Metropolis Hastings algorithm, we find no significant violation of duality, then model selection criteria favor flat scenarios even though a slight curvature is not totally ruled out. For the $H_0$ tension we find a preference at $1$-$\sigma$ for $h^R_0=0.730\pm0.010$ from supernovae when dropping DESI data and for $h^P_0=0.674\pm 0.005$ from Planck when using DESI and gamma-ray bursts.
\end{abstract}

%\keywords{Keyword1; keyword2; keyword3.}

%\ccode{PACS numbers: 03.65.$-$w, 04.62.+v}

\section{Introduction}

The cosmic distance duality (CDD) relation relates the luminosity $d_L$ and angular-diameter $d_A$ distances as $d_L/d_A=(1+z)^2$ derived from the reciprocity theorem \cite{ellis2007}. Specifically, the theorem states that two observers at rest to each other in a static spacetime will see two objects subtend identical solid angles leading to a relation between area distances, i.e. $r^2_G=r^2_0(1+z)^2$ where $r_G$ is the area distance of the source, say a galaxy while $r_0$ is the area distance of the observer also labeled \emph{distance by apparent size}, i.e. $d_A$. The area distance of the galaxy, considering an emission of radiation, can be seen as directly observing $d_L$ leading to $d^2_L=r^2_G(1+z)^2$. Thus, considering the relation between the two area distances and since $r_0\equiv d_A$ one ends up with $d^2_L=r^2_0(1+z)^4=d^2_A(1+z)^4$. Departures from it may hint at new physics or systematic errors in distance measurements. Furthermore, recently it has been discussed that investigating the violation of the CDD can help shed light on the $H_0$ tension. In this context, the tension can be seen as a \emph{cosmic calibration tension}\cite{coscal} arising when the luminosity and angular-diameter distances are compared as in the CDD relation. To search for possible violation of the relation we define
\begin{equation}\label{CDDviol}
    \frac{d_L}{d_A(1+z)^2}=\eta(z),
\end{equation}
where $\eta(z)$ is a function that quantifies deviations from duality. 

We consider five parameterizations of $\eta(z)$ to assess possible deviations from the CDD relation, specifically: i) a Taylor expansion (TE) to map epochs at $z\leq 1$ and ii) a power-law (PL) parameterization. Starting from ii) to overcome convergence issues arising from using Taylor we consider iii) a logarithmic (LOG) correction, iv) a Pad\'e series of order $(2;1)$ and v) a second order Chebyshev (T2) parameterization.

Then, we adopt model-independent and -dependent approaches. The first adopts B\'ezier polynomials to parameterize $H(z)$ while for the latter we consider the $\Lambda$CDM\footnote{Although the success of the concordance paradigm it suffers from the cosmological constant problem\cite{belfiglio2023} from which derive also the fine-tuning and coincidence problems. Thus, different models have been proposed, e.g. from thermodynamic models \cite{dunsby2016} or considering the dark energy fluid not as a cosmological constant but as a repulsive effect \cite{luongoquevedo2014}. Moreover, also the nature of dark matter remains an open issue afflicting the concordance paradigm, alternatives to it for example have been proposed in Ref. \cite{luongo2025}.} and the $\omega_0\omega_1$CDM models widely debated since DESI data releases\cite{desi}. 

Even though a violation of the CDD relation is not related to the background cosmological model we decide to explore the model-dependent scenario to see how the cosmological parameters would react to a possible deviation from $\eta(z)= 1$.

We use probes at low- and intermediate/high-$z$ to obtain constraints on all parameters by performing a Monte-Carlo Markov chain (MCMC) analysis using the Metropolis-Hastings algorithm. We start with \emph{Analysis 1} which adopts observational Hubble data (OHD), angular-diameter distances of galaxy clusters obtained through the Sunyaev-Zeldovich (SZ) effect, type Ia supernovae (SNeIa) and the first DESI data release (DR1). Then, \emph{Analysis 2} is the same as \emph{Analysis 1} but without DR1-DESI. Afterwards, adopting the Akaike information criterion (AIC) and the deviance information criterion (DIC) we find a preference for a spatially flat universe when using B\'ezier and for the flat $\Lambda$CDM model in the model-dependent case. However, in both cases a slight curvature is not totally ruled out\cite{alfano2026cosmic}.

Then, when using GRB data we introduce the second DESI data release (DR2) adopting OHD+SZ+SNeIa+DR2-DESI using the $E_p-E_{iso}$ (or \emph{Amati}) correlation for \emph{Analysis A} or the $L_0-E_p-T$ (or \emph{Combo}) correlation for \emph{Analysis C}\cite{alfano2025investigating}.

Our computations show no evident violation of the CDD relation. Concerning $h_0$, our results agree at $1$-$\sigma$ with $h^R_0=0.730\pm 0.010$ \cite{2022ApJ...934L...7R} when the DR1-DESI is dropped, i.e. \emph{Analysis 2} while it is in agreement with $h^P_0=0.674\pm 0.005$ \cite{planck} when DR2-DESI and GRBs are considered, i.e. \emph{Analyses A-C}.

This work is organized as follows. In Sect. \ref{CDDparam} we present the parameterizations while Sect. \ref{modes} focuses on the model-independent and model-dependent approaches. Sect. \ref{cprobes} deals with the cosmic probes employed in our analyses whose results from the MCMC and model selection criteria are in Sect. \ref{numr}. Sect. \ref{conc} focuses on the conclusions.

\section{Cosmic distance duality parameterizations}\label{CDDparam}

Departures from the CDD relation could hint at new physics or possible systematic errors in the measurement of the distances. For the latter point, it can be linked to the $H_0$ tension in terms of a \emph{cosmic calibration tension}\cite{coscal}. We consider five different frameworks to test the relation, starting from a Taylor series of $\eta(z)$ around $z\simeq 0$
\begin{equation}\label{taylor}
    \eta(z)=1+\eta_0z.
\end{equation}
This first parameterization is adopted in order to have both distances as infinitesimal of the same order at redshift $z=0$ and to quantify possible departures from the duality relation at low-$z$. However, the Taylor series suffers from a \emph{convergence problem}. To address and overcome this issue, we consider other four $\eta(z)$ functions starting from a power-law parameterization
\begin{equation}\label{transp}
    \eta(z)=(1+z)^{\eta_0}.
\end{equation}
First proposed in Ref. \cite{Avgoustidis:2009ai} this parameterization also addresses departures from the \emph{cosmic transparency} facing directly a possible violation of the conservation of the photon number. To further investigate this last point the temperature of the cosmic microwave background (CMB) radiation is considered in terms of $\eta_0$.
Then, from Eq. \eqref{transp} we consider that for $|\eta_0|\ll 1$ we have a LOG parameterization
\begin{equation}\label{log}
 \eta(z)=\exp\left[\eta_0\ln(1+z)\right]\simeq1+\eta_0\ln(1+z).
\end{equation}
The logarithmic correction pictured in Eq. \eqref{log} ensures a weaker evolution at both low- ($z\ll 1$) and high-$z$ regimes ($z\gg 1$). However, using Eq. \eqref{log} does not fully heal the convergence issue. In light of this, we adopt the Pad\'e and Chebyshev polynomials. Always considering Eq. \eqref{transp} a $(1;2)$ Pad\'e series reads\cite{alfano2026cosmic, alfano2025investigating}
\begin{equation}\label{pade}
  \eta^{(1;2)}(z)\simeq\frac{6+2z(2+\eta_0)}{6+4z(1-\eta_0)-z^2(1-\eta_0)\eta_0}.
\end{equation}
The adoption of $\eta(z)$ parameterized using a Pad\'e series ensure that Eq. \eqref{pade} does not diverge within the range of data we adopt for our computations.
Our last parameterization is derived through a second order Chebyshev polynomial\cite{alfano2026cosmic}
\begin{equation}\label{cheby}
 \eta^{(2)}(z)\simeq 2(1+z)^{2\eta_0}-1.
\end{equation}
The adoption of Chebyshev polynomials minimize the uncertainties on $\eta_0$. This addresses and heal a problem with Pad\'e approximants which, on the other hand, exhibit at higher redshifts an increase in the error bars since they are derived from a Taylor expansion.

In all parameterizations discussed here the CDD relation is resumed when $\eta_0=0$.

\section{Model-independent and model-dependent approaches}\label{modes}

Here we discuss the two approaches used in this work. Starting from the model-independent approach, it is based on the reconstruction of the Hubble rate $H(z)$ via B\'ezier polynomials. The advantage of this methodology lies in its non-linear monotonic growing trend when we stop at the second order of the B\'ezier curve showing in this way an agreement with the behavior of $H(z)$
\begin{equation}\label{Hzbez}
    H_2(y)=100 \alpha_0\left[(1-y)^2+2\frac{\alpha_1}{\alpha_0}y(1-y)+\frac{\alpha_2}{\alpha_0}y^2\right]\ \text{km/s/Mpc},
\end{equation}
where $\alpha_i=\{\alpha_0,\alpha_1,\alpha_2\}$ are B\'ezier coefficients with $\alpha_0\equiv h_0$, i.e. the reduced Hubble constant and $0\leq y\equiv z/z_M\leq 1$ with $z_M$ the maximum redshift of the used sample.

Another model-independent reconstruction of $H(z)$ can be done through the \emph{cosmographic approach} \cite{luongo2015, Luongo:2013rba, luongomuccino2020, luongoquevedo2012} based on expanding the Hubble rate in a Taylor series up to a selected order. However, due to convergence issues arising at $z\geq 1$ when Taylor is adopt one can reconstruct the Hubble rate using, e.g. Pad\'e approximants, Chebyshev polynomials or auxiliary variables. Although both the B\'ezier and cosmographic approaches reconstruct $H(z)$ without imposing {\it a priori} cosmological model some differences subsist. Primarily, while cosmography gives rise to cosmographic parameters encoding, e.g. the acceleration or deceleration of the universe through the parameter $q_0$ the B\'ezier curve only coefficient that reproduces a cosmological quantity is $\alpha_0\equiv h_0$. Furthermore, Eq. \eqref{Hzbez} does not have a convergence issue unlike Taylor-based cosmography. \footnote{Regarding this, the reader is referred to Fig. 2 of Ref. \protect\cite{Alfano:2023uff}.}

For the model-dependent approach we consider $H(z)$ with the contribution of matter $\Omega_m$, curvature $\Omega_k$ and dark energy $\Omega_{de}=1-\Omega_m-\Omega_k$ as
\begin{equation}\label{hrate}
   H(z)=H_0\sqrt{\Omega_m(1+z)^3+\Omega_k(1+z)^2+\Omega_{de}f(z)},   
\end{equation}
where the evolution of the dark energy fluid is depicted through $f(z)$ which varies according to what cosmological model we choose to test. Here, $f(z)=1$ for the $\Lambda$CDM model while for the $\omega_0\omega_1$CDM scenario $f(z)=(1+z)^{3(1+\omega_0+\omega_1)}\exp\left(-\frac{3\omega_1z}{1+z}\right)$.

\section{Cosmic probes}\label{cprobes}

This section is dedicated to the cosmic probes adopted in our analyses.

\begin{itemize}
    \item [-] {\bf OHD}. Model independent measurements of the Hubble rate found by considering passively evolving red galaxies from which values of $H(z)$ are derived considering the age difference between pairs of galaxies near each other but at different $z$. Specifically, the values are derived from $H(z)=-(\Delta z/\Delta t)(1+z)^{-1}$ with $\Delta t$ and $\Delta z$ the age and redshift difference.
    \item [-] {\bf SZ data}. When using this data set, we need to be careful since the derived distances depend on the validity of the CDD relation\cite{uzan} and thus $d_A(z)$ needs to be written as\cite{uzan, holanda}
\begin{equation}
    d_A^{SZ}(z)=d^\star_A(z)\eta^2(z),
\end{equation}
where $d^\star_A(z)$ is
\begin{equation}\label{dA}
    d^\star_A(z)=\frac{cH^{-1}_0}{(1+z)\sqrt{|\Omega_k|}}S_k\left[\int^z_0\frac{\sqrt{|\Omega_k|}H_0}{H^\star(z^\prime)}dz^\prime\right],
\end{equation}
where $H^\star(z)$ varies according to if we are considering Eq. \eqref{Hzbez} or Eq. \eqref{hrate} and $S_k(x)$ varies according to if we are considering a flat or non-flat scenario.
\item [-] {\bf SNeIa}. We consider the Pantheon catalog of SNeIa consisting in 1048 sources. For this sample their distance moduli take the form
\begin{equation}
    \mu_S(z)=25+5\log[\eta(z)(1+z)^2d^\star_A(z)],
\end{equation}
with $\eta(z)$ depends on the parameterization and $d^\star_A(z)$ is defined in Eq. \eqref{dA}.
\item [-] {\bf DESI-BAO}. We consider measurements from the two DESI data releases \cite{desi} fixing the baryon drag epoch to $r_d=(147.09\pm 0.26)\ \text{Mpc}$ \cite{planck}. Assuming a violation of the CDD relation the distances are
\begin{equation}\label{DESIdis}
   \frac{d^\star_M(z)}{r_d}=\frac{\eta(z)d^\star_A(z)(1+z)}{r_d},\quad  \frac{d^\star_H(z)}{r_d}=\frac{c}{r_dH^\star(z)},\quad \frac{d^\star_V(z)}{r_d}=\frac{[zd^\star_H(z)d^{2\star}_M(z)]^{1/3}}{r_d}.
\end{equation}
where Eqs. \eqref{DESIdis} vary according to the treatment we decide to adopt.

    \item [-] {\bf $E_p-E_{iso}$ correlation}. It relates the rest-frame peak energy $E_p$ and the isotropic radiated energy of the burst $E_{iso}$
    \begin{equation}\label{amati}
        \log\left(\frac{E_p}{\text{keV}}\right)=b-a\left[52-\log\left(\frac{E_{iso}}{\text{erg}}\right)\right],
    \end{equation}
    where $a$ is the slope and $b$ is the intercept of the correlation. When Eq. \eqref{amati} is adopted to investigate possible deviations from the CDD relation $E_{iso}$ is 
    \begin{equation}\label{eiso}
        E_{iso}=4\pi d^2_2(z)\eta^2(z)(1+z)^3S_b,
    \end{equation}
    where $S_b$ is the bolometric fluence while $d_2(z)$ is the angular-diameter distance when Eq. \eqref{Hzbez} is used.

\item [-] {\bf $L_0-E_p-T$ correlation}. It combines the prompt emission quantity $E_p$ with the plateau luminosity $L_0$ and the rest-frame duration of the plateau phase of the X-Ray afterglow emission $T$
    \begin{equation}\label{combo}
        \log\left(\frac{L_0}{\text{erg/s}}\right)=b+a\log\left(\frac{E_p}{\text{keV}}\right)-\log\left(\frac{T}{\text{s}}\right),
    \end{equation}
    with $a$ as the slope and $b$ as the intercept. When considering Eq. \eqref{CDDviol} the luminosity $L_0$ takes the following form
    \begin{equation}\label{l0}
        L_0=4\pi d^2_2(z)\eta^2(z)(1+z)^4F_0,
    \end{equation}
    where also here $d_2(z)$ is the angular-diameter distance when Eq. \eqref{Hzbez} is used. 
\end{itemize}

\section{Numerical results}\label{numr}

We consider four different analyses dealing with combinations of the different probes discussed in this work. Specifically, \emph{Analyses 1-2} focus on the impact of the DR1-DESI sample on all the parameters involved, including $\eta_0$ using the parameterizations depicted in Eq. \eqref{taylor} and Eqs. \eqref{log}-\eqref{cheby}. For these analyses we find the best-fit parameters shown in Tabs. \ref{tab:bfBezier}-\ref{tab:bfCosmo} by maximizing the following log-likelihoods\cite{alfano2026cosmic}
\begin{align*}
    &\textbf{Analysis 1:}\ \ln\mathcal{L}=\ln\mathcal{L}_X+\ln\mathcal{L}_{DR1},\\
    &\textbf{Analysis 2:}\ \ln\mathcal{L}=\ln\mathcal{L}_X.
\end{align*}
Afterwards, \emph{Analyses A-C} focus only on the model-independent approach and deal with how different GRB correlations affects $\eta_0$ considering the $\eta(z)$ parameterizations in Eqs. \eqref{transp}-\eqref{pade} together with employing the DR2-DESI sample. Here, the best-fit parameters are in Tab. \ref{tab:bfGRB} and are found by maximizing the log-likelihoods\cite{alfano2025investigating}
\begin{align*}
  \textbf{Analysis A:}\ \ln\mathcal{L}=\ln\mathcal{L}_X+\ln\mathcal{L}_{DR2}+\ln\mathcal{L}_A,\\
  \textbf{Analysis C:}\  \ln\mathcal{L}=\ln\mathcal{L}_X+\ln\mathcal{L}_{DR2}+\ln\mathcal{L}_C.
\end{align*}
In both \emph{Analyses 1-2} and \emph{Analyses A-C} $\ln\mathcal{L}_X=\ln\mathcal{L}_O+\ln\mathcal{L}_S+\ln\mathcal{L}_{SZ}$.

\subsection{Analysis 1}

We focus our attention on \emph{Analysis 1} where we consider the impact of DESI-DR1. A more complete discussion on it is found in Ref. \cite{alfano2026cosmic}. 

\begin{itemize}
    \item [-] {\bf Model-independent approach}. 
    Starting from the flat scenario $\alpha_0\equiv h_0$ agrees only at $2$-$\sigma$ with $h^P_0=0.674\pm 0.005$ \cite{planck}. For the non-flat case, there is no compatibility between our $h_0$ and Planck $h^P_0 = 0.636^{+0.021}_{-0.023}$  \cite{planck}. Comparing our $\Omega_k$ with $\Omega_k=-0.011^{+0.013}_{-0.012}$ \cite{planck} it agrees
    at $1$-$\sigma$ for all four parameterizations.
    
      For $\eta_0$, no substantial deviation from $\eta_0=0$ is found.
    
    \item [-] {\bf Model-dependent approach}. For this case, $h_0$ tends to agree at $2$-$\sigma$ \emph{only} in the flat scenario with $h^P_0$ with both the $\Lambda$CDM and $\omega_0\omega_1$CDM models. Then, $\Omega_k$ is compatible at $1$-$\sigma$ with Planck for both scenarios.

    Focusing on the $\omega_0$ and $\omega_1$ parameters of the $\omega_0\omega_1$CDM model, in the flat case $\omega_0$ and $\omega_1$ agree \emph{only} at $2$-$\sigma$ with the predictions of the concordance paradigm, namely $\omega_0 = -1$ and $\omega_1 = 0$. When $\Omega_k\neq 0$ $\omega_0$ agrees with $\omega_0=-1$ at $1$-$\sigma$ while $\omega_1$ is compatible with zero only at $2$-$\sigma$.

    Concerning $\eta_0$, as for the model-independent approach, there is no meaningful violation of the CDD relation.
\end{itemize}

\subsection{Analysis 2}

We discuss in \emph{Analysis 2} what happens when the DESI-DR1 sample is dropped. A more complete discussion on it is found in Ref. \cite{alfano2026cosmic}. 

\begin{itemize}
    \item [-] {\bf Model-independent approach}. We find an agreement at $1$-$\sigma$ between our $h_0$ and $h^R_0$ \cite{2022ApJ...934L...7R}. Afterwards, when the curvature is considered, no compatibility between our $h_0$ and $h^P_0$ is found while our $\Omega_k$ agrees at $1$-$\sigma$ with Planck's $\Omega_k$ \cite{planck} for all parameterizations.

    Concerning $\eta_0$, also in \emph{Analysis 2} there is no considerable departures from zero in all the involved parameterizations.
    
    \item [-] {\bf Model-dependent approach}. In the flat scenario, our $h_0$ agrees with $h^R_0$ \cite{2022ApJ...934L...7R} at $1$-$\sigma$ for both the concordance paradigm and $\omega_0\omega_1$CDM scenario. In the curvature case, for both models, $h_0$ does not agree with $h^P_0$\cite{planck} while the curvature parameter agrees at $1$-$\sigma$ with $\Omega_k$ from Planck\cite{planck}.
    
    Regarding $\omega_0$ and $\omega_1$, in both cases of a null and non-null curvature parameter, our values agrees at $1$-$\sigma$ with $\omega_0=-1$ and at $2$-$\sigma$ with $\omega_1=0$. 

   Again, also in the model-dependent approach we find no significant evidence for a violation of the CDD relation, i.e. our $\eta_0$ agrees with $\eta_0\approx 0$.
\end{itemize}

\begin{table*}
\scriptsize
\centering
\setlength{\tabcolsep}{1.2em}
\renewcommand{\arraystretch}{1.1}
 \resizebox{\columnwidth}{!}{
\begin{tabular}{lcccccc}
\hline
$\alpha_0\equiv h_0$  &  $\alpha_1$ & $\alpha_2$ & $\Omega_k$ & $\eta_0$ & $\Delta\text{AIC}$ & $\Delta\text{DIC}$ \\
\hline\hline
\multicolumn{7}{c}{{\bf Analysis 1}}\\
\hline\hline
\multicolumn{7}{c}{{TE}}\\
\hline
$0.692^{+0.008(0.016)}_{-0.008(0.016)}$ & $1.113^{+0.025(0.058)}_{-0.028(0.060)}$ & $2.395^{+0.044(0.089)}_{-0.042(0.082)}$ & $0$ & $0.009^{+0.008(0.017)}_{-0.008(0.017)}$ & $0$ & $0$\\
$0.690^{+0.010(0.017)}_{-0.006(0.014)}$ & $1.115^{+0.026(0.063)}_{-0.032(0.067)}$ & $2.394^{+0.040(0.093)}_{-0.039(0.081)}$ & $-0.059^{+0.365(0.786)}_{-0.253(0.490)}$ & $0.016^{+0.030(0.060)}_{-0.045(0.084)}$ & $2$ & $2$\\
\hline
\multicolumn{7}{c}{{ LOG}}\\
\hline

$0.693^{+0.007(0.015)}_{-0.009(0.017)}$ & $1.112^{+0.031(0.064)}_{-0.028(0.060)}$ & $2.395^{+0.040(0.087)}_{-0.043(0.084)}$ &   $0$ & $0.015^{+0.014(0.028)}_{-0.015(0.030)}$ & $0$ & $0$ \\
$0.691^{+0.008(0.017)}_{-0.008(0.015)}$ & $1.115^{+0.029(0.064)}_{-0.033(0.069)}$ & 
$2.394^{+0.043(0.091)}_{-0.041(0.084)}$ & $-0.002^{+0.194(0.387)}_{-0.139(0.280)}$ & $0.017^{+0.026(0.057)}_{-0.036(0.076)}$ & $2$ & $2$\\
\hline
\multicolumn{7}{c}{{ P(1;2)}}\\
\hline
$0.691^{+0.008(0.017)}_{-0.007(0.016)}$ & $1.117^{+0.026(0.058)}_{-0.033(0.064)}$ & $2.388^{+0.049(0.093)}_{-0.036(0.079)}$ & $0$ & $0.015^{+0.012(0.026)}_{-0.014(0.029)}$ & $0$ & $0$\\
$0.691^{+0.009(0.016)}_{-0.006(0.015)}$ & $1.117^{+0.027(0.063)}_{-0.034(0.069)}$ & $2.393^{+0.045(0.088)}_{-0.043(0.084)}$ & $-0.008^{+0.209(0.396)}_{-0.141(0.319)}$ & $0.017^{+0.025(0.059)}_{-0.040(0.078)}$ & $2$ & $4$\\
\hline
\multicolumn{7}{c}{T2}\\
\hline
$0.691^{+0.009(0.017)}_{-0.007(0.015)}$ & $1.115^{+0.029(0.062)}_{-0.031(0.063)}$ & $2.394^{+0.040(0.085)}_{-0.043(0.082)}$ & $0$ & $0.003^{+0.004(0.008)}_{-0.003(0.007)}$ & $0$ & $0$\\
$0.692^{+0.006(0.015)}_{-0.008(0.016)}$ & $1.117^{+0.027(0.061)}_{-0.036(0.069)}$ & $2.390^{+0.044(0.090)}_{-0.036(0.076)}$ & $+0.031^{+0.151(0.342)}_{-0.184(0.335)}$ & $0.002^{+0.007(0.016)}_{-0.008(0.017)}$ & $2$ & $5$\\
 \hline\hline
 \multicolumn{7}{c}{{\bf Analysis 2}}\\
\hline\hline
\multicolumn{7}{c}{{ TE}}\\
\hline
$0.705^{+0.021(0.044)}_{-0.021(0.041)}$ & $0.965^{+0.063(0.129)}_{-0.051(0.113)}$ & $2.141^{+0.134(0.297)}_{-0.173(0.337)}$ & $0$ & $-0.023^{+0.027(0.059)}_{-0.029(0.058)}$ & $0$ & $0$\\
$0.704^{+0.023(0.046)}_{-0.018(0.039)}$ & $0.909^{+0.096(0.181)}_{-0.054(0.109)}$ & $2.239^{+0.144(0.317)}_{-0.209(0.409)}$ & $+0.922^{+0.460(1.171)}_{-0.957(1.659)}$ & $-0.119^{+0.109(0.190)}_{-0.045(0.100)}$ & $1$ & $1$\\
\hline
\multicolumn{7}{c}{{ LOG}}\\
\hline
$0.708^{+0.021(0.045)}_{-0.022(0.043)}$ & $0.946^{+0.074(0.150)}_{-0.063(0.135)}$ & $2.142^{+0.152(0.319)}_{-0.147(0.303)}$ & $0$ & $-0.048^{+0.054(0.105)}_{-0.042(0.088)}$ & $0$ & $0$\\
$0.710^{+0.018(0.044)}_{-0.024(0.044)}$ & $0.929^{+0.083(0.179)}_{-0.087(0.175)}$ & $2.173^{+0.201(0.407)}_{-0.152(0.378)}$ & $+0.233^{+0.392(0.880)}_{-0.448(0.800)}$ & $-0.080^{+0.090(0.173)}_{-0.086(0.175)}$ & $2$ & $2$\\
\hline
\multicolumn{7}{c}{{ P(1;2)}}\\
\hline
$0.707^{+0.023(0.048)}_{-0.022(0.043)}$ & $0.938^{+0.084(0.161)}_{-0.060(0.135)}$ & $2.157^{+0.128(0.295)}_{-0.170(0.329)}$ & $0$ & $-0.049^{+0.053(0.103)}_{-0.047(0.096)}$ & $0$ & $0$\\
$0.706^{+0.023(0.049)}_{-0.019(0.042)}$ & $0.934^{+0.081(0.173)}_{-0.082(0.183)}$ & $2.159^{+0.223(0.426)}_{-0.144(0.349)}$ & $+0.128^{+0.453(0.923)}_{-0.340(0.731)}$ & $-0.070^{+0.063(0.163)}_{-0.086(0.178)}$ & $2$ & $2$\\
\hline
\multicolumn{7}{c}{T2}\\
\hline
$0.704^{+0.026(0.050)}_{-0.017(0.039)}$ & $0.957^{+0.068(0.139)}_{-0.080(0.152)}$ & $2.118^{+0.177(0.344)}_{-0.125(0.272)}$ & $0$ & $-0.010^{+0.011(0.023)}_{-0.013(0.026)}$ & $0$ & $0$\\
$0.708^{+0.021(0.044)}_{-0.022(0.044)}$ & $0.925^{+0.094(0.182)}_{-0.084(0.180)}$ & $2.223^{+0.162(0.350)}_{-0.190(0.415)}$ & $+0.130^{+0.456(0.897)}_{-0.340(0.744)}$ & $-0.017^{+0.016(0.039)}_{-0.022(0.045)}$ & $2$ & $2$\\
 \hline
\end{tabular}}
\caption{Best-fit B\'ezier coefficients and $\eta_0$ in the flat(non-flat) scenario with attached errors at $1$-$\sigma$ ($2$-$\sigma$). The table, taken from Ref.  \cite{alfano2026cosmic}, also depicts the preferred scenario when the AIC and DIC are adopted.}
\label{tab:bfBezier}
\end{table*}

\begin{table}
\scriptsize
\centering
\setlength{\tabcolsep}{1.2em}
\renewcommand{\arraystretch}{1.1}
\resizebox{\columnwidth}{!}{
\begin{tabular}{lcccccccc}
\hline
& $h_0$ & $\Omega_m$ & $\Omega_k$ & $\omega_0$ & $\omega_1$ & $\eta_0$ & $\Delta\text{AIC}$ & $\Delta\text{DIC}$\\
\hline\hline
\multicolumn{9}{c}{\textbf{Analysis 1}}\\
\hline\hline
\multicolumn{9}{c}{TE}\\
\hline
$\Lambda$CDM & $0.694^{+0.007(0.015)}_{-0.008(0.016)}$ & $0.299^{+0.013(0.027)}_{-0.012(0.024)}$ & $0$ & $-1$ & $0$ & $0.007^{+0.008(0.015)}_{-0.009(0.016)}$ & $0$ & $0$\\
$\Lambda$CDM+$\Omega_k$ & $0.694^{+0.007(0.015)}_{-0.009(0.017)}$ & $0.290^{+0.027(0.055)}_{-0.025(0.051)}$ & $+0.023^{+0.076(0.144)}_{-0.066(0.139)}$ & $-1$ & $0$ & $0.005^{+0.009(0.018)}_{-0.008(0.017)}$ & $2$ & $3$\\
$\omega_0\omega_1$CDM & $0.689^{+0.009(0.018)}_{-0.006(0.016)}$ & $0.321^{+0.012(0.030)}_{-0.025(0.093)}$ & $0$ & $-0.872^{+0.095(0.229)}_{-0.113(0.207)}$ & $-0.961^{+0.921(1.848)}_{-0.588(1.400)}$ & $0.008^{+0.010(0.018)}_{-0.007(0.015)}$ & $3$ & $4$\\
$\omega_0\omega_1$CDM+$\Omega_k$ &
$0.691^{+0.008(0.016)}_{-0.008(0.016)}$ & $0.342^{+0.048(0.126)}_{-0.075(0.187)}$ & $-0.104^{+0.270(0.444)}_{-0.191(0.597)}$ & $-0.822^{+0.134(0.269)}_{-0.201(0.548)}$ & $-0.625^{+0.409(1.914)}_{-0.705(1.889)}$ & $0.023^{+0.027(0.080)}_{-0.036(0.052)}$ & $5$ & $7$
\\
\hline
\multicolumn{9}{c}{LOG}\\
\hline
$\Lambda$CDM & $0.694^{+0.007(0.015)}_{-0.008(0.015)}$ & $0.298^{+0.012(0.025)}_{-0.012(0.024)}$ & $0$ & $-1$ & $0$ & $0.009^{+0.012(0.025)}_{-0.012(0.025)}$ & $0$ & $0$\\
$\Lambda$CDM+$\Omega_k$ & $0.692^{+0.008(0.016)}_{-0.008(0.016)}$ & $0.291^{+0.022(0.047)}_{-0.028(0.053)}$ & $+0.023^{+0.079(0.151)}_{-0.055(0.124)}$ & $-1$ & $0$ & $0.006^{+0.013(0.027)}_{-0.013(0.025)}$ & $2$ & $2$\\
$\omega_0\omega_1$CDM & $0.690^{+0.008(0.017)}_{-0.007(0.016)}$ & $0.317^{+0.014(0.033)}_{-0.020(0.065)}$ & $0$ & $-0.872^{+0.114(0.247)}_{-0.105(0.204)}$ & $-0.813^{+0.634(1.439)}_{-0.689(1.514)}$ & $0.013^{+0.016(0.033)}_{-0.010(0.027)}$ & $3$ & $3$\\
$\omega_0\omega_1$CDM+$\Omega_k$ & $0.690^{+0.006(0.015)}_{-0.004(0.014)}$ & $0.333^{+0.005(0.091)}_{-0.072(0.126)}$ & $-0.051^{+0.241(0.414)}_{-0.060(0.258)}$ & $-0.788^{+0.009(0.203)}_{-0.246(0.590)}$ & $-0.980^{+0.611(1.649)}_{-0.788(3.374)}$ & $0.028^{+0.010(0.048)}_{-0.051(0.087)}$ & $5$ & $12$\\
\hline
\multicolumn{9}{c}{P(1;2)}\\
\hline
$\Lambda$CDM & $0.694^{+0.008(0.015)}_{-0.007(0.015)}$ & $0.298^{+0.013(0.027)}_{-0.012(0.024)}$ & $0$ & $-1$ & $0$ & $0.008^{+0.013(0.025)}_{-0.011(0.024)}$ & $0$ & $0$\\
$\Lambda$CDM+$\Omega_k$ & $0.692^{+0.008(0.017)}_{-0.007(0.016)}$ & $0.283^{+0.029(0.055)}_{-0.020(0.044)}$ & $+0.047^{+0.053(0.123)}_{-0.083(0.148)}$ & $-1$ & $0$ & $0.003^{+0.015(0.029)}_{-0.010(0.023)}$ & $2$ & $2$\\
$\omega_0\omega_1$CDM & $0.690^{+0.008(0.017)}_{-0.006(0.016)}$ & $0.315^{+0.018(0.034)}_{-0.016(0.055)}$ & $0$ & $-0.864^{+0.107(0.244)}_{-0.099(0.215)}$ & $-0.760^{+0.492(1.370)}_{-0.723(1.714)}$ & $0.017^{+0.012(0.026)}_{-0.016(0.032)}$ & $3$ & $4$\\
$\omega_0\omega_1$CDM+$\Omega_k$ & $0.692^{+0.006(0.014)}_{-0.012(0.017)}$ & $0.335^{+0.062(0.112)}_{-0.078(0.224)}$ & $-0.082^{+0.302(0.461)}_{-0.207(0.379)}$ & $-0.829^{+0.153(0.281)}_{-0.236(0.668)}$ & $-0.657^{+0.232(1.984)}_{-1.512(4.280)}$ & $0.027^{+0.045(0.080)}_{-0.048(0.095)}$ & $5$ & $10$\\
\hline
\multicolumn{9}{c}{T2}\\
\hline
$\Lambda$CDM & $0.693^{+0.008(0.016)}_{-0.007(0.015)}$ &  $0.299^{+0.013(0.026)}_{-0.013(0.024)}$ & $0$ & $-1$ & $0$ & $0.002^{+0.003(0.006)}_{-0.003(0.006)}$ & $0$ & $0$\\
 $\Lambda$CDM+$\Omega_k$ & $0.692^{+0.008(0.016)}_{-0.007(0.015)}$ & $0.288^{+0.024(0.049)}_{-0.024(0.048)}$ & $+0.033^{+0.064(0.139)}_{-0.063(0.131)}$ & $-1$ & $0$ & $0.001^{+0.003(0.007)}_{-0.003(0.006)}$ & $2$ & $3$\\
$\omega_0\omega_1$CDM &
$0.692^{+0.005(0.015)}_{-0.009(0.018)}$ &  $0.315^{+0.018(0.034)}_{-0.017(0.070)}$ & $0$ & $-0.853^{+0.109(0.225)}_{-0.116(0.233)}$ & $-0.896^{+0.719(1.608)}_{-0.643(1.478)}$ & $0.004^{+0.003(0.007)}_{-0.004(0.008)}$ & $3$ & $5$\\
$\omega_0\omega_1$CDM+$\Omega_k$ &
$0.692^{+0.003(0.014)}_{-0.009(0.016)}$ &  $0.300^{+0.066(0.133)}_{-0.048(0.104)}$ & $+0.060^{+0.144(0.324)}_{-0.205(0.477)}$ & $-0.945^{+0.227(0.386)}_{-0.164(0.463)}$ & $-0.802^{+0.609(1.372)}_{-0.767(3.310)}$ & $0.001^{+0.001(0.025)}_{-0.007(0.016)}$ & $5$ & $5$\\
\hline\hline
\multicolumn{9}{c}{\textbf{Analysis 2}}\\
\hline\hline
\multicolumn{9}{c}{TE}\\
\hline
$\Lambda$CDM & $0.702^{+0.021(0.043)}_{-0.020(0.040)}$ & $0.277^{+0.043(0.090)}_{-0.038(0.076)}$ & $0$ & $-1$ & $0$ & $-0.017^{+0.029(0.058)}_{-0.026(0.053)}$ & $0$ & $0$\\
$\Lambda$CDM+$\Omega_k$ & $0.704^{+0.020(0.043)}_{-0.020(0.041)}$ & $0.307^{+0.085(0.170)}_{-0.084(0.159)}$ & $-0.075^{+0.170(0.346)}_{-0.146(0.311)}$ & $-1$ & $0$ & $-0.013^{+0.028(0.059)}_{-0.028(0.055)}$ & $2$ & $2$\\
$\omega_0\omega_1$CDM & $0.705^{+0.014(0.038)}_{-0.025(0.045)}$ & $0.349^{+0.058(0.110)}_{-0.041(0.147)}$ & $0$ & $-0.894^{+0.212(0.478)}_{-0.195(0.371)}$ & $-2.937^{+2.081(3.976)}_{-2.274(5.191)}$ & $-0.036^{+0.038(0.068)}_{-0.015(0.047)}$ & $2$ & $3$\\
$\omega_0\omega_1$CDM+$\Omega_k$ & $0.699^{+0.016(0.038)}_{-0.018(0.041)}$ & $0.212^{+0.217(0.469)}_{-0.107(0.187)}$ & $+0.345^{+0.208(0.472)}_{-0.472(1.305)}$ & $-1.101^{+0.519(0.797)}_{-0.240(0.679)}$ & $-4.935^{+2.850(7.013)}_{-5.953(9.377)}$ & $-0.059^{+0.053(0.155)}_{-0.025(0.065)}$ & $4$ & $5$

\\
\hline
\multicolumn{9}{c}{LOG}\\
\hline
$\Lambda$CDM & $0.701^{+0.021(0.042)}_{-0.021(0.040)}$ & $0.280^{+0.043(0.090)}_{-0.039(0.074)}$ & $0$ & $-1$ & $0$ & $-0.020^{+0.040(0.081)}_{-0.037(0.075)}$ & $0$ & $0$\\
$\Lambda$CDM+$\Omega_k$ & $0.706^{+0.020(0.045)}_{-0.022(0.042)}$ & $0.311^{+0.076(0.150)}_{-0.064(0.141)}$ & $-0.088^{+0.148(0.325)}_{-0.150(0.298)}$ & $-1$ & $0$ & $-0.021^{+0.037(0.082)}_{-0.037(0.074)}$ & $2$ & $2$\\
$\omega_0\omega_1$CDM & $0.705^{+0.018(0.043)}_{-0.025(0.046)}$ & $0.363^{+0.040(0.092)}_{-0.057(0.140)}$ & $0$ & $-0.949^{+0.227(0.510)}_{-0.189(0.398)}$ & $-3.287^{+2.363(4.477)}_{-1.933(5.580)}$ & $-0.050^{+0.042(0.096)}_{-0.037(0.087)}$ & $2$ & $2$\\
$\omega_0\omega_1$CDM+$\Omega_k$ & $0.704^{+0.019(0.045)}_{-0.025(0.046)}$ & $0.263^{+0.176(0.291)}_{-0.106(0.201)}$ & $+0.245^{+0.194(0.364)}_{-0.455(0.689)}$ & $-0.994^{+0.316(0.791)}_{-0.347(0.899)}$ & $-5.453^{+3.471(6.340)}_{-6.125(9.483)}$ & $-0.072^{+0.065(0.151)}_{-0.057(0.110)}$ & $4$ & $4$\\
\hline
\multicolumn{9}{c}{P(1;2)}\\
\hline
$\Lambda$CDM & $0.701^{+0.020(0.042)}_{-0.020(0.040)}$ & $0.279^{+0.043(0.088)}_{-0.038(0.076)}$ & $0$ & $-1$ & $0$ & $-0.021^{+0.040(0.080)}_{-0.038(0.077)}$ & $0$ & $0$\\
$\Lambda$CDM+$\Omega_k$ & $0.708^{+0.019(0.042)}_{-0.024(0.044)}$ & $0.313^{+0.072(0.153)}_{-0.072(0.144)}$ & $-0.100^{+0.171(0.342)}_{-0.127(0.290)}$ & $-1$ & $0$ & $-0.024^{+0.041(0.082)}_{-0.034(0.074)}$ & $2$ & $2$\\
$\omega_0\omega_1$CDM & $0.697^{+0.028(0.054)}_{-0.015(0.036)}$ & $0.371^{+0.036(0.085)}_{-0.056(0.150)}$ & $0$ & $-0.937^{+0.230(0.509)}_{-0.171(0.392)}$ & $-2.955^{+2.005(4.029)}_{-2.472(5.514)}$ & $-0.036^{+0.034(0.082)}_{-0.063(0.105)}$ & $2$ & $2$\\
$\omega_0\omega_1$CDM+$\Omega_k$ & $0.706^{+0.012(0.037)}_{-0.022(0.047)}$ & $0.241^{+0.143(0.337)}_{-0.068(0.179)}$ & $+0.267^{+0.164(0.332)}_{-0.349(0.753)}$ & $-1.027^{+0.297(0.768)}_{-0.271(0.693)}$ & $-4.934^{+3.475(5.916)}_{-4.439(9.962)}$ & $-0.079^{+0.039(0.133)}_{-0.042(0.104)}$ & $4$ & $4$\\
\hline
\multicolumn{9}{c}{T2}\\
\hline
$\Lambda$CDM & $0.701^{+0.021(0.041)}_{-0.020(0.040)}$ &  $0.282^{+0.042(0.089)}_{-0.041(0.078)}$ & $0$ & $-1$ & $0$ & $-0.005^{+0.009(0.020)}_{-0.010(0.019)}$ & $0$ & $0$\\
 $\Lambda$CDM+$\Omega_k$ & $0.704^{+0.023(0.047)}_{-0.019(0.041)}$ &  $0.319^{+0.068(0.147)}_{-0.077(0.149)}$ & $-0.095^{+0.168(0.341)}_{-0.139(0.300)}$ & $-1$ & $0$ & $-0.005^{+0.009(0.019)}_{-0.010(0.020)}$ & $2$ & $2$\\
$\omega_0\omega_1$CDM & $0.702^{+0.020(0.045)}_{-0.021(0.044)}$ &  $0.363^{+0.043(0.093)}_{-0.057(0.159)}$ & $0$ & $-0.961^{+0.233(0.537)}_{-0.168(0.368)}$ & $-2.928^{+2.134(4.092)}_{-2.541(5.926)}$ & $-0.012^{+0.012(0.025)}_{-0.011(0.022)}$ & $2$ & $2$
\\
$\omega_0\omega_1$CDM+$\Omega_k$ & $0.704^{+0.014(0.044)}_{-0.021(0.045)}$ &  $0.271^{+0.123(0.266)}_{-0.099(0.263)}$ & $+0.214^{+0.232(0.407)}_{-0.287(0.675)}$ & $-1.069^{+0.343(0.901)}_{-0.245(0.724)}$ & $-4.831^{+3.063(7.141)}_{-6.494(10.13)}$ & $-0.022^{+0.015(0.037)}_{-0.011(0.024)}$ & $4$ & $3$
\\
\hline
\caption{Best-fit cosmological parameters and $\eta_0$ for the flat(non-flat) $\Lambda$CDM and $\omega_0\omega_1$CDM models with attached errors at $1$-$\sigma$($2$-$\sigma$). The table, taken from Ref. \cite{alfano2026cosmic}, also depicts the preferred scenario when the AIC and DIC are adopted. }
\label{tab:bfCosmo}
\end{tabular}
}
\end{table}

\subsection{Analysis A and Analysis C}

Here we present the results for both \emph{Analysis A}, i.e. when we consider the  $E_p-E_{iso}$ and \emph{Analysis C} when the consider the  $L_0-E_p-T$ correlation function. For both analyses we include the DESI-DR2. A more complete discussion on it is found in Ref. \cite{alfano2025investigating}. 

\begin{itemize}
    \item [-] {\bf Model-independent approach}. Regardless of the GRB correlation, for all parameterizations involved our $h_0$ agrees with $h^P_0=0.674\pm 0.005$ \cite{planck} at $1$-$\sigma$, while no compatibility was found with $h^R_0=0.730\pm 0.010$ \cite{2022ApJ...934L...7R}. 

    Then, $\eta_0$ agrees with zero at $1$-$\sigma$ for all three parameterizations, thus observing no violation of the CDD relation.
\end{itemize}

\begin{table*}
\scriptsize
\centering
\setlength{\tabcolsep}{0.6em}
\renewcommand{\arraystretch}{1.}
\resizebox{\columnwidth}{!}{
\begin{tabular}{ccccccc}
\hline
\multicolumn{7}{c}{{\bf Analysis A}}\\
\hline
 $a$ & $b$ & $\sigma$ & $\alpha_0\equiv h_0$ & $\alpha_1$ & $\alpha_2$ & $\eta_0$ \\
\hline
\multicolumn{7}{c}{PL}\\
\cline{1-7}
$0.854^{+0.099(0.172)}_{-0.087(0.136)}$ & $1.618^{+0.131(0.203)}_{-0.129(0.227)}$ & $0.327^{+0.066(0.119)}_{-0.046(0.074)}$ & $0.683^{+0.010(0.017)}_{-0.011(0.016)}$ & $1.054^{+0.027(0.046)}_{-0.025(0.043)}$ & $2.002^{+0.024(0.041)}_{-0.030(0.046)}$ & $0.003^{+0.016(0.027)}_{-0.014(0.024)}$\\
\hline
\multicolumn{7}{c}{LOG}\\
\cline{1-7}
$0.856^{+0.090(0.163)}_{-0.085(0.133)}$ & $1.621^{+0.124(0.190)}_{-0.125(0.235)}$ & $0.337^{+0.049(0.103)}_{-0.056(0.080)}$ & $0.682^{+0.012(0.019)}_{-0.009(0.016)}$ & $1.060^{+0.022(0.038)}_{-0.032(0.048)}$ & $1.996^{+0.029(0.046)}_{-0.023(0.039)}$ & $0.004^{+0.017(0.025)}_{-0.015(0.023)}$\\
\hline
\multicolumn{7}{c}{P(1,2)}\\      
\cline{1-7}
$0.856^{+0.099(0.178)}_{-0.087(0.135)}$ & $1.625^{+0.121(0.189)}_{-0.144(0.245)}$ & $0.332^{+0.062(0.111)}_{-0.052(0.077)}$ & $0.683^{+0.010(0.017)}_{-0.011(0.018)}$ & $1.052^{+0.027(0.046)}_{-0.023(0.040)}$ & $2.000^{+0.026(0.043)}_{-0.028(0.044)}$ & $0.005^{+0.014(0.025)}_{-0.015(0.025)}$\\
\hline
\multicolumn{7}{c}{{\bf Analysis C}}\\
\hline
 $a$ & $b$ & $\sigma$ & $\alpha_0\equiv h_0$ & $\alpha_1$ & $\alpha_2$ & $\eta_0$ \\
\hline
\multicolumn{7}{c}{PL}\\
\cline{1-7}
$0.821^{+0.105(0.172)}_{-0.080(0.159)}$ & $49.702^{+0.215(0.422)}_{-0.269(0.456)}$ & $0.368^{+0.044(0.072)}_{-0.036(0.060)}$ & $0.681^{+0.011(0.018)}_{-0.009(0.015)}$ & $1.057^{+0.022(0.039)}_{-0.029(0.046)}$ & $1.998^{+0.026(0.042)}_{-0.026(0.043)}$ & $0.001^{+0.016(0.025)}_{-0.014(0.025)}$\\
\hline
\multicolumn{7}{c}{LOG}\\
\cline{1-7}
$0.844^{+0.081(0.154)}_{-0.117(0.190)}$ & $49.655^{+0.300(0.484)}_{-0.220(0.412)}$ & $0.367^{+0.042(0.069)}_{-0.034(0.056)}$ & $0.682^{+0.010(0.017)}_{-0.011(0.017)}$ & $1.056^{+0.024(0.041)}_{-0.027(0.046)}$ & $1.999^{+0.025(0.042)}_{-0.026(0.044)}$ & $0.004^{+0.015(0.024)}_{-0.017(0.027)}$\\
\hline
\multicolumn{7}{c}{P(1,2)}\\      
\cline{1-7}
$0.828^{+0.092(0.167)}_{-0.095(0.171)}$ & $49.671^{+0.262(0.462)}_{-0.221(0.421)}$ & $0.372^{+0.038(0.067)}_{-0.040(0.061)}$ & $0.681^{+0.011(0.018)}_{-0.010(0.016)}$ & $1.053^{+0.027(0.043)}_{-0.025(0.044)}$ & $1.999^{+0.026(0.045)}_{-0.028(0.041)}$ & $0.002^{+0.015(0.025)}_{-0.015(0.025)}$\\
\hline
\end{tabular}
}
\caption{Best-fit correlation parameters, B\'ezier coefficients and $\eta_0$, respectively for both \emph{Analysis A} and \emph{Analysis C}. This table is taken from Ref. \cite{alfano2025investigating}.}
\label{tab:bfGRB}
\end{table*}

\section{Outlooks and perspectives}\label{conc}

In this work, we investigate possible violations of the CDD relation adopting five parameterizations of $\eta(z)$. We consider model-independent (through B\'ezier) and -dependent (through the $\Lambda$CDM and $\omega_0\omega_1$CDM models) approaches combining cosmic probes mapping low- and intermediate/high-$z$ epochs. We also check how the parameters react to potential deviations from Eq. \eqref{CDDviol} and the preferred scenario through the AIC and DIC. Furthermore, we investigate departures from duality in light of recent works who have suggested that deviations from $\eta(z)=1$ can be related to the $H_0$ tension as a \emph{cosmic calibration tension}\cite{coscal}.

We perform MCMC analyses adopting the Metropolis-Hastings algorithm considering as primary probes OHD, SZ data and SNeIa. Our analyses are then enforced with the DR1-DESI, DR2-DESI and GRBs to probe regions at higher $z$.

In \emph{Analysis 1} and \emph{Analysis 2} no substantial violation of the CDD relation is observed and through the AIC and DIC we see a preference towards a flat scenario for the model-independent case and for the flat $\Lambda$CDM model even in \emph{Analysis 1} although a slight curvature is not totally ruled out. Focusing on the $H_0$ tension we find a preference at $1$-$\sigma$ with $h^R_0=0.730\pm 0.01$ \cite{2022ApJ...934L...7R} when DESI data are dropped.

When GRB data are considered we only adopt the model-independent approach. Also in this case no violation of the CDD relation is observed since $\eta_0\approx 0$ at $1$-$\sigma$ in every parameterization. Regarding $h_0$, including GRB correlations together with the DR2-DESI gives $1$-$\sigma$ agreement with $h^P_0$ and none with $h^R_0$ \cite{2022ApJ...934L...7R}.

Future works will focus on reducing errors on the inferred parameters considering additional cosmographic analyses together with complement our analyses with other high-$z$ probes such as standard sirens to investigate further the Hubble tension.

\end{document}